\documentclass{mpe_report}

\usepackage{psfig,graphicx,epsfig}
\usepackage{color}
\usepackage{amsmath,amssymb,epic,eepic,array}

\begin{document}

\pagenumbering{arabic}
\setcounter{page}{104}

\renewcommand{\FirstPageOfPaper }{104}\renewcommand{\LastPageOfPaper }{107}%%
%% This is MPE-Report_example.tex
%% LaTeX2e example style file for the contributed talks and posters presented
%% during the 363rd Heraeus Seminar on Neutron Stars and Pulsars, held in 
%% Bad Honnef, May 14.-19. 2006.
%% 
%% This file needs the LaTeX2e class file he_symp.cls  
%%
\newcommand{\Ell}{E_\parallel}      % E_\parallel
\newcommand{\rhoGJ}{\rho_{{\rm GJ}}}  % Goldreich-Julian charge density
\newcommand{\rlc}{\varpi_{\rm LC}} % light cylinder radius
\newcommand{\inc}{\alpha_{\rm i}}  % inclination angle of oblique rotator
\newcommand{\omgp}{\omega_{\rm p}} % 
\newcommand{\rhowSQR}{\rho_{\rm w}^2}
\newcommand{\etaICe}{\eta_{\rm IC}^{\rm e}}
\newcommand{\etaICg}{\eta_{\rm IC}^\gamma}

% -----------------------------------------------------------------------------
%\documentclass{mpe_report}
%\usepackage{psfig}
% -----------------------------------------------------------------------------
%\def\R{~ROSAT}
%\def\RAS{\R all sky survey}
% -----------------------------------------------------------------------------
%\begin{document}

\title{Particle Accelerator in Pulsar Magnetospheres:
 A Hybrid Solution of Inner and Outer Gap Models}
\author{K. Hirotani}
\institute{ASIAA/National Tsing Hua University - TIARA, 
           PO Box 23-141, Taipei, Taiwan
           {\it Email: hirotani@tiara.sinica.edu.tw} \quad
           Postal address: TIARA, Department of Physics, 
           National Tsing Hua University,
           101, Sec. 2, Kuang Fu Rd.,Hsinchu, Taiwan 300}
\titlerunning{Particle Accelerator in Pulsar Magnetospheres}
\maketitle

\begin{abstract}
A self-consistent electrodynamics of a particle accelerator 
in a rotating neutron-star magnetosphere is investigated
on the two-dimensional poloidal plane.
Solving the Poisson equation for the electrostatic potential
together with the Boltzmann equations for 
electrons, positrons and gamma-rays,
it is demonstrated that the created current density increases to be
super-Goldreich-Julian if the trans-field thickness of the gap becomes
thick enough.
This new solution exists from the neutron-star surface
to the outer magnetosphere
with a small-amplitude positive acceleration field in the inner part,
which works to extract ions from the stellar surface 
as a space-charge-limited flow.
The acceleration field is highly unscreened in the outer magnetosphere,
in the same manner as in traditional outer-gap models.
\end{abstract}

\section{Introduction}

The Energetic Gamma Ray Experiment Telescope (EGRET) 
aboard the Compton Gamma Ray Observatory 
has detected pulsed signals from at least six rotation-powered pulsars.
Since interpreting $\gamma$-rays should be less ambiguous
compared with reprocessed, non-thermal X-rays,
the $\gamma$-ray pulsations observed from these objects
are particularly important as a direct signature of 
basic non-thermal processes in pulsar magnetospheres.

Attempts to model the particle accelerator % pulsed $\gamma$-ray emissions
have traditionally concentrated on two scenarios:
Polar-cap models with emission altitudes within
several neutron star radii over a polar cap surface 
(Harding, Tademaru, \& Esposito 1978; Daugherty \& Harding 1982, 1996;
 Sturner, Dermer, \& Michel 1995),
and outer-gap models with acceleration occurring in the open
zone located near the light cylinder
(Cheng, Ho, \& Ruderman 1986a,b;
 Chiang \& Romani 1992, 1994; Romani and Yadigaroglu 1995).
Both models predict that electrons and positrons are
accelerated in a charge depletion region, a potential gap,
by the electric field along the magnetic field lines
to radiate high-energy $\gamma$-rays via the curvature 
and inverse-Compton processes.

The outer gap models 
have been successful in explaining the observed light curves,
particularly in reproducing the wide separation of the two
peaks commonly observed from $\gamma$-ray pulsars
(Kanbach 1999; Thompson 2001).
In these models,
they consider that the gap extends from the null surface,
where the Goldreich-Julian charge density vanishes,
to the light cylinder,
beyond which the velocity of a co-rotating plasma would 
exceed the velocity of light,
adopting the vacuum solution of the Poisson equation for
the electrostatic potential.

However, it was analytically demonstrated by 
Hirotani, Harding, and Shibata (2003, HHS03)
that the gap inner boundary shifts towards the star
as the created current increases
and at last touch the star if the created current exceeds
the Goldreich-Julian (GJ) value at the surface.
Therefore, 
to understand the particle accelerator,
which extends from the stellar surface to the outer magnetosphere,
we have to merge the outer-gap and polar-cap models,
which have been separately considered so far.

In traditional polar-cap models,
the energetics and pair cascade spectrum have had success
in reproducing the observations.
However, the predicted beam size of radiation emitted 
near the stellar surface
is too small to produce the wide pulse profiles that are observed
from high-energy pulsars.
Seeking the possibility of a wide hollow cone of high-energy radiation
due to the flaring of field lines,
Arons (1983) first examined the particle acceleration
at the high altitudes along the last open field line.
This type of accelerator, or the slot gap, 
forms because the accelerating electric field, $\Ell$,
is screened at increasingly higher altitude 
as the magnetic colatitude
approaches the edge of the open field region
(Arons \& Scharlemann 1979).
Muslimov and Harding (2003, 2004a,b, hereafter MH03, MH04a,b)
extended this argument by including two new features:
acceleration due to space-time dragging, and
the additional decrease of $\Ell$ at the edge of the gap
due to the narrowness of the slot gap.
It is noteworthy that the polar-slot gap model proposed by MH04a,b
is an extension of the polar-cap model
into the outer magnetosphere,
assuming that the plasma flowing in the gap consists of only one sign of 
charges, which leads to a negative $\Ell$ 
when the rotation and magnetic moment vectors resides in the same hemisphere.
However, we should notice here 
that the electric current induced by the negative $\Ell$
contradicts with the global current patterns,
which is derived by the EMF exerted on the spinning neutron-star surface,
if the gap is located near the last-open field line.

On these grounds, we are motivated by the need to contrive 
an accelerator model that predicts a consistent current direction
with the global requirement. 
To this aim, it is straightforward to extend recent outer-gap models,
which predict opposite $\Ell$ to polar-cap models,
into the inner magnetosphere.
Extending the one-dimensional analysis along the field lines 
in several outer-gap models
(Hirotani and Shibata 1999a,~b,~c; HHS03), 
Takata, Shibata, and Hirotani (2004, hereafter TSH04)
and Takata et al. (2006, hereafter TSHC06) 
solved the Poisson equation for the electrostatic potential
on the two-dimensional poloidal plane,
and revealed that the gap inner boundary is located inside of the
null surface owing to the pair creation within the gap,
assuming that the particle motion immediately saturates
in the balance between electric and radiation-reaction forces.
In the present paper, we extend TSH04 and TSHC06
by solving the particle energy distribution explicitly,
and by considering a super-GJ current solution with ion emission
from the neutron star surface.
                 
\section{Basic Equations and Boundary Conditions}
\label{sec:basic}
We formulate the basic equations to describe the
particle accelerator, extending the method first proposed by 
Beskin et al. (1992) for black-hole magnetospheres.
The first kind equation we have to consider is the
Poisson equation for the non-corotational potential,
\begin{equation}
  -\frac{c^2}{\sqrt{-g}}
   \partial_\mu 
      \left( \frac{\sqrt{-g}}{\rhowSQR}
             g^{\mu\nu} g_{\varphi\varphi}
             \partial_\nu \Psi
      \right)
  = 4\pi(\rho-\rhoGJ),
  \label{eq:Poisson_2}
\end{equation}
where the general relativistic Goldreich-Julian charge density
is defined as
\begin{equation}
  \rhoGJ \equiv 
      \frac{c^2}{4\pi\sqrt{-g}}
      \partial_\mu \left[ \frac{\sqrt{-g}}{\rhowSQR}
                         g^{\mu\nu} g_{\varphi\varphi}
                         (\Omega-\omega) F_{\varphi\nu}
                 \right].
  \label{eq:def_GJ}
\end{equation}

The second kind equations we have to consider is the 
Boltzmann equations for particles.
Imposing a stationary condition $\partial_t +\Omega\partial_\phi=0$,
utilizing $\mbox{\boldmath$\nabla$}\cdot\mbox{\boldmath$B$}=0$,
and introducing dimensionless particle densities per unit magnetic flux tube
such that $n_\pm = N_\pm / (\Omega B/2\pi ce)$,
we obtain the following form of the particle Boltzmann equations, 
\begin{equation}
  c\cos\chi \frac{\partial n_\pm}{\partial s}
  +\frac{dp}   {dt}\frac{\partial n_\pm}{\partial p}
  +\frac{d\chi}{dt}\frac{\partial n_\pm}{\partial \chi}
  = S_\pm,
 \label{eq:BASIC_2}
\end{equation}
where $s$ denotes the distance along the magnetic field line,
($p$,$\chi$) the momentum and pitch angle of the particles.
The upper and lower signs correspond to the
positrons (with charge $q=+e$) and 
electrons ($q=-e$), respectively; 
$p \equiv \vert\mbox{\boldmath$p$}\vert$ and $ds/dt=c\cos\chi$,
\begin{equation}
  \frac{dp}{dt} \equiv q\Ell\cos\chi -\frac{P_{\rm SC}}{c}
  \label{eq:char1}
\end{equation}
\begin{equation}
  \frac{d\chi}{dt} \equiv -\frac{q\Ell\sin\chi}{p}
                         +c\frac{\partial(\ln B^{1/2})}{\partial s}
                          \sin\chi,
  \label{eq:char2}  
\end{equation}
For outward- (or inward-) migrating particles,
$\cos\chi>0$ (or $\cos\chi<0$).
Since we consider relativistic particles, 
we obtain $\Gamma=p/(m_{\rm e}c)$.
Using $n_\pm$, we can express $\rho_{\rm e}$ as
\begin{equation}
  \rho_{\rm e}
  = \frac{\Omega B}{2\pi c}
    \int\!\!\!\!\int \left[ n_+(s,\theta_\ast,\Gamma,\chi)
                           -n_-(s,\theta_\ast,\Gamma,\chi)
                     \right]
    d\Gamma d\chi.
  \label{eq:def_rhoe}
\end{equation}
The synchro-curvature radiation-reaction force, $P_{\rm SC}/c$, 
is given by Cheng and Zhang (1996).
Collision terms $S_\pm$ represents the particle appearing and disappearing
rate from ($p$,$\chi$) and are given by the
inverse-Compton scatterings and one-photon and two-photon pair creation
processes.

% Let us briefly mention the electric current per magnetic flux tube.
% With projected velocities, $c\cos\chi$, along the field lines,
% electric current density in units of $\Omega B / (2\pi)$ is given by
% \begin{equation}
%   j_{\rm gap}(s,\theta_\ast)
%   = j_{\rm e}(s,\theta_\ast)+j_{\rm ion}(\theta_\ast),
%   \label{eq:jgap}
% \end{equation}
% where
% \begin{equation}
%   j_{\rm e} \equiv \!\!\int\!\!\!\!\int (n_- +n_+) \cos\chi \, dpd\chi;
% \end{equation}
% $j_{\rm ion}$ denotes the current density carried by the ions
% emitted from the stellar surface.
% Since most of the particles have relativistic velocities 
% projected along the magnetic field lines at each point,  
% $j_{\rm gap}$ is kept virtually constant for $s$.

The third kind equations we have to consider is the 
Boltzmann equation for $\gamma$-rays.
Imposing the stationary condition, we obtain
\begin{equation}
  \left( c\frac{k^\varphi}{\vert\mbox{\boldmath$k$}\vert}
        -\Omega \right)
   \frac{\partial g}{\partial\bar{\varphi}}
 +c\frac{k^r}{\vert\mbox{\boldmath$k$}\vert}
   \frac{\partial g}{\partial r}
 +c\frac{k^\theta}{\vert\mbox{\boldmath$k$}\vert}
   \frac{\partial g}{\partial \theta}
 = S_\gamma,
  \label{eq:BASIC_3}
\end{equation}
where $\bar{\varphi}=\varphi-\Omega t$.
To compute $k^i$, 
we have to solve the photon propagation in the curved spacetime.
For the computation of $k^i$ and $S_\gamma$, see Hirotani (2006).

In order to solve the set of partial differential 
equations~(\ref{eq:Poisson_2}), (\ref{eq:BASIC_2}), and
(\ref{eq:BASIC_3})
for $\Psi$, $n_\pm$, and $g$,
we must impose appropriate boundary conditions.
We assume that the gap {\it lower} boundary,
$\theta_\ast=\theta_\ast^{\rm max}$,
coincides with the last open field line.
Moreover, we assume that the {\it upper} boundary coincides with a specific
magnetic field line
and parameterize this field line with 
$\theta_\ast=\theta_\ast^{\rm min}$.
Determining the upper boundary from physical consideration
is a subtle issue,
which is beyond the scope of the present paper.
Therefore, we treat $\theta_\ast^{\rm min}$ as a free parameter.
We measure the trans-field thickness of the gap with
\begin{equation}
  h_{\rm m} \equiv \frac{\theta_\ast^{\rm max}-\theta_\ast^{\rm min}}
                        {\theta_\ast^{\rm max}}.
  \label{eq:def_hm}
\end{equation}
If $h_{\rm m}=1.0$, it means that the gap exists along 
all the open field lines. 
On the other hand, if $h_{\rm m}\ll 1$, 
the gap becomes transversely thin. 
To describe the trans-field structure,
we introduce the fractional height as
\begin{equation}
  h \equiv \frac{\theta_\ast^{\rm max}-\theta_\ast}
                {\theta_\ast^{\rm max}}.
\end{equation}
Thus, the lower and upper boundaries
are given by $h=0$ and $h=h_{\rm m}$, respectively.
The {\it inner} boundary is assumed to be located at the 
neutron star surface.
For the {\it outer} boundary,
we solve the Poisson equation to the light cylinder
(see Hirotani 2006 for details).

First, to solve the elliptic-type equation~(\ref{eq:Poisson_2}),
we impose $\Psi=0$ on the lower, upper, and inner boundaries.
At the outer boundary, we impose $\partial\Psi/\partial s=0$.
Generally speaking, 
the solved $\Ell=-(\partial\Psi/\partial s)_{s \rightarrow 0}$
under these boundary conditions
does not vanish at the stellar surface.
For a super-GJ current density in the sense that
$\rho_{\rm e}-\rhoGJ<0$ holds at the stellar surface,
equation~(\ref{eq:Poisson_2}) gives a positive electric field near the star.
In this case, we assume that ions
are emitted from the stellar surface so that the additional positive
charge in the thin non-relativistic region may bring $\Ell$ to zero.

Secondly, to solve the hyperbolic-type equations~(\ref{eq:BASIC_2})
and (\ref{eq:BASIC_3}),
we assume that neither positrons nor $\gamma$-rays are injected
across the inner boundary; thus, we impose 
\begin{equation}
  n_+(s^{\rm in},\theta_\ast,\Gamma,\chi) = 0, \quad
  g(s^{\rm in},\theta_\ast,E_\gamma,\theta_\gamma) = 0 
  \label{eq:BD-2}
\end{equation}
for arbitrary $\theta_\ast$, $\Gamma$, $0<\chi<\pi/2$, $E_\gamma$,
and $\cos(\theta_\gamma-\theta_{\rm B})>0$,
where $\theta_{\rm B}$ designates the outward magnetic field direction.
In the same manner, at the outer boundary, we impose
\begin{equation}
  n_-(s^{\rm out},\theta_\ast,\Gamma,\chi) = 0, \quad
  g(s^{\rm out},\theta_\ast,E_\gamma,\theta_\gamma) = 0 
  \label{eq:BD-4}
\end{equation}
for arbitrary $\theta_\ast$, $\Gamma$, $\pi/2<\chi<\pi$, $E_\gamma$,
and $\cos(\theta_\gamma-\theta_{\rm B})<0$.

\section{Application to the Crab Pulsar}
\label{sec:appl}
We apply the scheme to the Crab pulsar.
We present the created current density, $j_{\rm e}$,
as a function of $h$, for the five different trans-field thickness,
$h_{\rm m}=0.047$ (dotted), $0.048$ (solid), $0.060$ (dashed),
$0.100$ (dash-dotted), and $0.160$ (dash-dot-dot-dotted).
The thin dashed line represents 
$\vert \rhoGJ/(\Omega B/2\pi c) \vert_{s=0}$;
if $j_{\rm e}$ appears below (or above) this line,
the created current is sub- (or super-) GJ along the field line
specified by $h$.
For $h_{\rm m}=0.047$, the solution (dotted curve) is sub-GJ 
along all the field lines;
thus, screening due to the discharge is negligible 
as shown by the dotted curve in figure~\ref{fig:c70b},
which depicts $\Ell(s,h)$ at the central height $h=h_{\rm m}/2$. 
As $h_{\rm m}$ increases, the solution becomes super-GJ from the
higher latitudes, as indicated by the solid,
dashed, dash-dotted, and dash-dot-dot-dotted curves.
As a result, $\Ell$ is screened significantly with increasing $h_{\rm m}$,
as demonstrated by figure~\ref{fig:c70b}.
This screening of $\Ell$ has a negative feed back effect 
in the sense that $j_{\rm e}$ is regulated below unity.

\begin{figure}
\centerline{\psfig{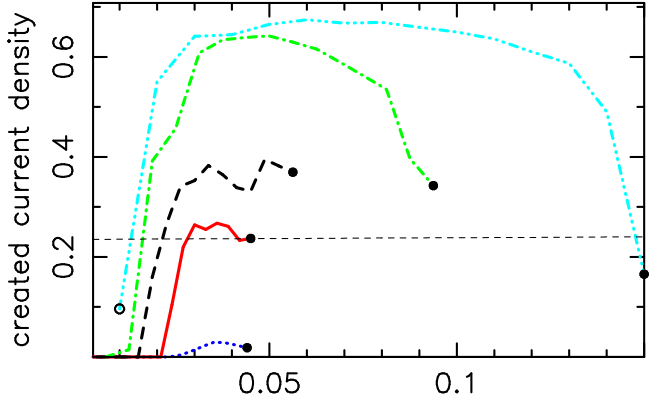}}
\caption{Created current density $j_{\rm e}$ (in unit of $\Omega B/2\pi$)
as a function of the transfield thickness $h$ for five different
trans-field heights.
\label{fig:c70b}}
\end{figure}

\begin{figure}
\centerline{\psfig{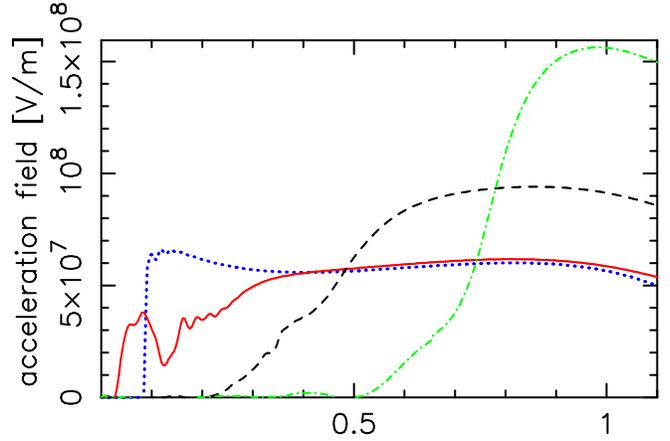}}
\caption{Field-aligned electric field at $h=h_{\rm m}/2$ 
as a function of $s/\rlc$.
The dotted, solid, dashed, and dash-dotted curves corresponds to
$h_{\rm}=0.047$, $0.048$, $0.060$, and $0.100$, respectively.
\label{fig:c70b}}
\end{figure}

To understand the screening mechanism, 
it is helpful to examine the Poisson equation~(\ref{eq:Poisson_2}).
In the transversely thin limit, it gives
\begin{equation}
  \Psi
  \approx \frac{\Omega B_\ast}{c}
          \frac{2\pi c}{\Omega B}(\rho-\rhoGJ),
          r_\ast^2 (\theta_\ast-\theta_\ast^{\rm min})
                   (\theta_\ast^{\rm max}-\theta_\ast).
  \label{eq:CHR}
\end{equation}
We thus find that
$\Ell \equiv -\partial\Psi/\partial s$ is approximately proportional to
$-\partial(\rho/B -\rhoGJ/B)/\partial s$. 
It is, therefore,
important to examine the two-dimensional distribution of 
$\rho/B$ and $\rhoGJ/B$ to understand $\Ell(s,h)$ behavior.

\begin{figure}
\centerline{\psfig{file=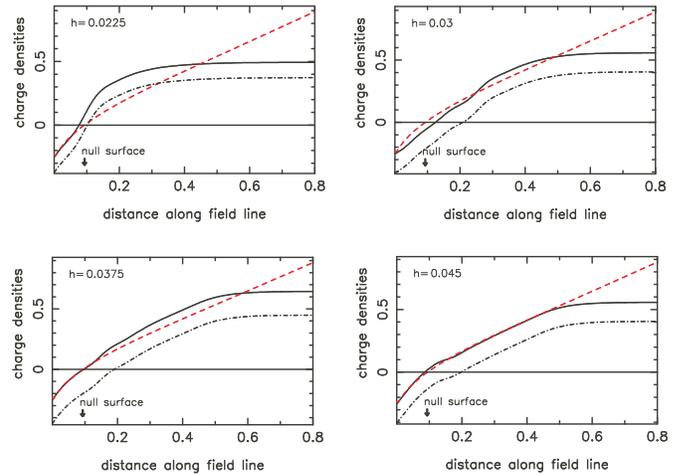,width=8.8cm,clip=}}
\caption{
Total (solid), created (dash-dotted), and Goldreich-Julian (dashed)
charge densities in $\Omega B(s,h) /(2\pi c)$ unit,
for $\inc=70^\circ$ and $h_{\rm m}=0.060$ 
at nine transfield heights, $h$.
If there is an ion emission from the stellar surface,
the total charge density deviates from the created one.
\label{fig:C70nine}
}
\end{figure}

In figure~\ref{fig:C70nine}, 
we present
$\rho/(\Omega B/2\pi c)$,  
$\rho_{\rm e}/(\Omega B/2\pi c)$, and 
$\rhoGJ/(\Omega B/2\pi c)$,
as the solid, dash-dotted, and dashed curves,
at nine discrete magnetic latitudes 
at $h=(3/8)h_{\rm m}$, $h_{\rm m}/2$, $(5/8)h_{\rm m}$, $(3/4)h_{\rm m}$,
for the same parameters as figure~\ref{fig:c70b}.
If there is a cold-field ion emission from the star,
the total charge density (solid curve) deviates from
the created charge density (dash-dotted one).
The current becomes super-GJ for $h \ge (6/16)h_{\rm m}=0.0225$.
Along the field lines with super-GJ current, 
$\rho_{\rm e}-\rhoGJ$ becomes negative close to the star. 
This inevitably leads to a positive $\Ell$,
which extracts ions from the stellar surface.

In the outer region, $\rho/B$ levels off 
in $s>0.5\rlc$ for $h>h_{\rm m}/2$.
Since $\rhoGJ/B$ becomes approximately a linear function of $s$,
$\Ell$ remains nearly constant in $s>0.5\rlc$,
in the same manner as in traditional outer-gap model.
In the inner region, on the other hand,
inward-directed $\gamma$-rays propagate into the convex side 
due to the field line curvature,
increasing particle density with $h$.
A two-dimensional effect in the Poisson equation becomes important 
in the higher altitudes ($0.1\rlc<s<0.3\rlc$)
along the higher-latitude field lines($h \ge 0.0225$).
Outside of the null surface, $s>0.09\rlc$,
there is a negative $\rho_{\rm eff}$ in the sub-GJ current region
($h \le 0.015$).
This negative $\rho_{\rm eff}$ 
works to prevent $\Ell$ from vanishing in the higher latitudes,
where pair creation is copious.
However, the created pairs discharge until $\Ell$ vanishes,
resulting in a larger gradient of $\rho$
than that of $\rhoGJ$ 
in the intermediate latitudes in 
$0.0225 \le h \le 0.0262$.
In the upper half region ($0.03 \le h < h_{\rm m}=0.06$),
$\partial\rho/\partial s$ does not have to be greater 
than $\partial\rhoGJ/\partial s$, 
in order to screen $\Ell$.

In short, the gap has a hybrid structure:
The lower latitudes (with small $h$) are nearly vacuum 
having sub-GJ current densities
and the inner boundary is located slightly inside of the null surface,
because $\Psi>0$ region will be
filled with the electrons emitted from the stellar surface.
The higher latitudes, on the other hand, are non-vacuum
having super-GJ current densities,
and the inner boundary is located at the stellar surface,
extracting ions at the rate such that their non-relativistic column density
at the stellar surface cancels the strong $\Ell$ 
induced by the negative $\rho-\rhoGJ$ of relativistic electrons, positrons,
and ions.
The created pairs discharge such that
$\Ell$ virtually vanishes in the region where pair creation is copious.
Thus, in the intermediate latitudes between the sub-GJ and super-GJ regions,
$\partial\rho/\partial s > \rhoGJ/\partial s$ holds. 

Even though the inner-most region of the gap is inactive, 
general relativistic effect 
(space-time dragging effect, in this case) is important to determine the
ion emission rate from the stellar surface.
For example, at $h=h_{\rm m}/2$ for $h_{\rm m}=0.600$
(i.e., the top right panel in fig.~\ref{fig:C70nine}),
$j_{\rm ion}$ is $69$\% greater than what would be obtained in
the Newtonian limit, 
$\rhoGJ=-\mbox{\boldmath$\Omega$}\cdot\mbox{\boldmath$B$}/2\pi c$.
This is because the reduced $\vert\rhoGJ\vert$ near the star 
(about $15$\% less than the Newtonian value)
enhances the positive $\Ell$,
which has to be canceled by a greater ion emission
(compared to the Newtonian value).
The current, $j_{\rm ion}$, 
is adjusted so that $\vert \rho_{\rm eff} \vert$
may balance with the trans-field derivative of $\Psi$ near the star.
The resultant $\vert \rho_{\rm eff} \vert$
becomes small compared to $\vert\rhoGJ\vert$,
in the same manner as in traditional polar-cap models,
which has a negative $\Ell$ with electron emission from the star.
Although the non-relativistic ions have a large positive charge 
density very close to the star (within $10$~cm from the surface), 
it cannot be resolved in figure~\ref{fig:C70nine}.
Note that the present calculation is performed from the stellar surface
to the outer magnetosphere and does not contain a region with $\Ell<0$.
It follows that an accelerator having $\Ell<0$
(e.g., a polar-cap or a polar-slot-gap accelerator)
cannot exist along the magnetic field lines that have an
super-GJ current density created by the mechanism described in the
present paper.

\begin{acknowledgements}
The author is grateful to Drs.
J.~G.~Kirk, B.~Rudak, K.~S.~Cheng, and A.~K.~Harding for helpful suggestions.
This work is supported by the Theoretical Institute for
Advanced Research in Astrophysics (TIARA) operated under Academia Sinica
and the National Science Council Excellence Projects program in Taiwan
administered through grant number NSC 94-2752-M-007-001,
and partly by KBN through the grant 2P03D.004.24 to B.~Rudak,
which enabled the author to use the MEDUSA cluster at CAMK Toru\'n.
He gratefully acknowledges the support by the WE-Heraeus foundation.
\end{acknowledgements}
   
% Example list of References

%\end{document}

        \clearpage


\begin{thebibliography}{} 
\bibitem[Arons \& Scharlemann(1979)]{aron79} 
    Arons, J., Scharlemann, E. T. 1979, ApJ 231, 854
\bibitem[Cheng et al.(1986a)]{chen86a} 
    Cheng, K. S., Ho, C., \& Ruderman, M., 1986a
    ApJ, 300, 500
\bibitem[Cheng et al.(1986b)]{chen86b} 
    Cheng, K. S., Ho, C., \& Ruderman, M., 1986b
    ApJ, 300, 522
\bibitem[Chiang \& Romani(1992)]{chia92} 
    Chiang, J., \&  Romani, R. W. 1992,
    ApJ, 400, 629
\bibitem[Chiang \& Romani(1994)]{chia94} 
    Chiang, J., \& Romani, R. W. 1994,
    ApJ, 436, 754
\bibitem[Daugherty \& Harding(1982)]{Daug82} 
    Daugherty, J. K., \& Harding, A. K. 1982, 
    ApJ, 252, 337
\bibitem[Daugherty \& Harding(1996)]{Daug96} 
    Daugherty, J. K., \& Harding, A. K. 1996, 
    ApJ, 458, 278
\bibitem[Harding et al.(1978)]{hard78} 
    Harding, A. K., Tademaru, E., \& Esposito, L. S. 1978, 
    ApJ, 225, 226
\bibitem[Hirotani \& Shibata(1999a)]{hiro99a} 
    Hirotani, K. \& Shibata, S.,
    1999a, MNRAS 308, 54 %  (Paper I)
\bibitem[Hirotani \& Shibata(1999b)]{hiro99b} 
    Hirotani, K. \& Shibata, S.,
    1999b, MNRAS 308, 67 % (Paper II)
\bibitem[Hirotani \& Shibata(1999c)]{hiro99c} 
    Hirotani, K. \&  Shibata, S.,
    1999c, PASJ 51, 683   % (Paper III)
\bibitem[Hirotani et al.(2003)]{hiro03}
    Hirotani, K., Harding, A. K., \& Shibata, S.,
    2003, ApJ 591, 334 (HHS03)
\bibitem[Hirotani et al.(2006)]{hiro06}
    Hirotani, K.
    2006, Mod. Phys. Lett. A (Brief Review) 21, 1319
    (astro-ph/0606017).
\bibitem[Kanbach (1999)]{kanb99} 
    Kanbach, G. 1999,
    in proc. of the Third INTEGRAL Workshop,
    ed. Bazzaro, A.,
    Astrophys. Lett. Comm. 38, 17
\bibitem[Muslimov \& Harding(2003)]{musl03} 
    Muslimov, A. G., \& Harding, A. K., 2003,
    ApJ, 588, 430
\bibitem[Muslimov \& Harding(2004)]{musl04a} 
    Muslimov, A. G., \& Harding, A. K., 2004a,
    ApJ, 606, 1143 (MH04a)
\bibitem[Muslimov \& Harding(2004)]{musl04b} 
    Muslimov, A. G., \& Harding, A. K., 2004b,
    ApJ, 617, 471 (MH04b)
\bibitem[Romani \& Yadigaroglu(1995)]{roma95} 
    Romani, R. W., \& Yadigaroglu, I. A. 1995,
    ApJ 438, 314
\bibitem[Sturner et al.(1995)]{stur95} 
    Sturner, S. J., Dermer, C. D., \& Michel, F. C. 1995, 
    ApJ 445, 736
\bibitem[Takata et al.(2004)]{taka04b} 
    Takata, J., Shibata, S., \& Hirotani, K., 2004,
    MNRAS 354, 1120 (TSH04)
\bibitem[Takata et al.(2006)]{taka06} 
    Takata, J., Shibata, S., Hirotani, K., \& Chang, H.-K., 2006,
    MNRAS 366, 1310 (TSHC06)
\bibitem[Thompson et al.(2001)]{thom01} 
  Thompson, D. J. 2001, in AIP Conf. Proc. 558,
  High Energy Gamma-Ray Astronomy, 
  ed. A. Goldwurm et al. (New York: AIP), 103
\end{thebibliography}
\end{document}